\pdfoutput=1
\documentclass[11pt,twoside,a4paper,cmspaper,final,collab]{cms-tdr}
\def\svnVersion{f891c9d}\def\svnDate{2026/09/28}\def\cmsCernNoTag{CERN-EP-2025-028}\def\cmsCernDate{\today}\def\cmsMessage{Published in Physical Review Letters as \href{http://dx.doi.org/10.1103/qydl-rlp7}{\doi{10.1103/qydl-rlp7}.}}
\begin{document}\cmsNoteHeader{SUS-23-007}

\providecommand{\PGgst}{\HepParticle{\PGg}{}{\ast}\xspace}
\providecommand{\PA}{\HepParticle{A}{}{}\xspace}
\providecommand{\CP}{\ensuremath{C\!P}\xspace}
\newcommand{\mttot}{\ensuremath{m_{\mathrm{T}}^{\text{tot}}}\xspace}
\newcommand{\mt}{\ensuremath{m_{\mathrm{T}}}\xspace}
\newcommand{\ettt}{\ensuremath{\Pe\tauh\tauh\tauh}\xspace}
\newcommand{\eett}{\ensuremath{\Pe\Pe\tauh\tauh}\xspace}
\newcommand{\emtt}{\ensuremath{\Pe\PGm\tauh\tauh}\xspace}
\newcommand{\mttt}{\ensuremath{\PGm\tauh\tauh\tauh}\xspace}
\newcommand{\mmtt}{\ensuremath{\PGm\PGm\tauh\tauh}\xspace}
\newcommand{\ttt}{\ensuremath{\tauh\tauh\tauh}\xspace}
\newcommand{\tttt}{\ensuremath{\tauh\tauh\tauh\tauh}\xspace}
\newcommand{\mH}{\ensuremath{m_{\PH}}\xspace}
\newcommand{\mh}{\ensuremath{m_{\Ph}}\xspace}
\newcommand{\mA}{\ensuremath{m_{\PA}}\xspace}
\newcommand{\mHc}{\ensuremath{m_{\PH^{\pm}}}\xspace}
\newcommand{\mphi}{\ensuremath{m_{\PGf}}\xspace}
\newcommand{\Phobs}{\ensuremath{\mathrm{\PH_{125}}}\xspace}
\newcommand{\jtt}{\ensuremath{\text{jet}\to\tauh}\xspace}
\newcommand{\Wjets}{\ensuremath{\PW}{+}\,\text{jets}\xspace}
\newcommand{\Zjets}{\ensuremath{\PZ}{+}\,\text{jets}\xspace}
\newcommand{\ZLL}{\ensuremath{\PZ/\PGgst\to\Pell\Pell}\xspace}
\newcommand{\FF}{\ensuremath{F_{\mathrm{F}}}\xspace}
\newcommand{\dR}{\ensuremath{{\Delta}R}\xspace}
\providecommand{\HIGGSTOOLS}{\textsc{HiggsTools-1}\xspace}

\ifthenelse{\boolean{cms@external}}%
{
      \providecommand{\cmsLeft}{upper\xspace}
      \providecommand{\cmsRight}{lower\xspace}
      \newcommand{\figurewidthfeynman}{0.8\linewidth}
      \newcommand{\figurewidthpostfit}{0.8\linewidth}
      \newcommand{\figurewidthmodelindependent}{0.9\linewidth}
      \newcommand{\figurewidthmodeldependent}{0.9\linewidth}
      \newcommand{\figurewidthmodeldependenttwod}{0.9\linewidth}
      \newcommand{\clearpagecms}{}
}{
    \providecommand{\cmsLeft}{left\xspace}
    \providecommand{\cmsRight}{right\xspace}
    \newcommand{\figurewidthfeynman}{0.4\linewidth}
    \newcommand{\figurewidthpostfit}{0.49\linewidth}
    \newcommand{\figurewidthmodelindependent}{0.55\linewidth}
    \newcommand{\figurewidthmodeldependent}{0.55\linewidth}
    \newcommand{\figurewidthmodeldependenttwod}{0.6\linewidth}
    \newcommand{\clearpagecms}{\clearpage}
}

\cmsNoteHeader{SUS-23-007}
\title{Search for the Nonresonant Production of a Pair of Additional Higgs Bosons in the Type-X Two-Higgs-Doublet Model in Proton-Proton Collisions at \texorpdfstring{$\sqrt{s} = 13\TeV$}{sqrt(s) = 13 TeV}}

\hypersetup{
pdfauthor={CMS Collaboration},
pdftitle={Search for the nonresonant production of a pair of additional Higgs bosons in the Type-X two-Higgs-doublet model in proton-proton collisions at sqrt(s) = 13 TeV},
pdfsubject={CMS},
pdfkeywords={CMS, Higgs, BSM, muon g-2}}

\date{\today}

\abstract{
  A search is presented for the production of two additional Higgs bosons from an off-shell \PZ boson, where both additional particles decay to \PGt lepton pairs. The search is performed with a data sample collected with the CMS detector from proton-proton collisions at the LHC at $\sqrt{s} = 13\TeV$, corresponding to an integrated luminosity of 138\fbinv. No deviation from the standard model background is observed. Exclusion limits are set on the type-X two-Higgs-doublet model alignment scenario. These results exclude previously unexplored regions of parameter space and provide constraints complementary to those from earlier searches. 
}

\maketitle

The standard model of particle physics (SM) provides a successful theoretical framework for describing the fundamental particles and their interactions in nature.
Despite its success, the SM does not explain several observed phenomena, including the nature of dark matter, the origin of neutrino masses, and the matter--antimatter asymmetry of the universe.
Extensions of the SM Higgs sector provide a well-motivated framework for addressing some of these shortcomings, for example by introducing additional sources of \CP violation, modifying the electroweak phase transition, or providing dark matter candidates~\cite{Ivanov:2017dad}.
One of the simplest such extensions is the two-Higgs-doublet model (2HDM), which adds a second complex scalar doublet to the SM Higgs sector and arises naturally in several beyond-the-SM (BSM) theories~\cite{Branco:2011iw}.
The enlarged scalar sector predicts additional Higgs bosons that may be produced and detected at the LHC~\cite{Ilnicka:2018def}.

A \CP-conserving 2HDM predicts five Higgs bosons; two neutral scalars (\Ph and \PH), one neutral pseudoscalar (\PA), and two charged scalars ($\PSHpm$)~\cite{Branco:2011iw}.
There are four types of \CP-conserving 2HDMs that are dependent on which Higgs doublet couples to which group of fermions (up-type quarks, down-type quarks, and charged leptons).
These are named type-I, II, X (``lepton-specific''), and Y (``flipped'').
A renormalizable \CP-conserving 2HDM without tree-level flavor-changing neutral currents can be parametrized by the Higgs boson masses (\mh, \mH, \mA, and \mHc), the ratio of the vacuum expectation values of the two Higgs doublets (\tanb), the mixing angle that determines the composition of \Ph and \PH ($\alpha$), and the mass-squared term in the Higgs potential ($m_{12}^{2}$)~\cite{Branco:2011iw}.
The alignment limit matches the 2HDM to the observed Higgs boson (\Phobs) by dictating that $\cos(\beta-\alpha)=0$ for $\Ph=\Phobs$ (normal scenario) or $\sin(\beta-\alpha)=0$ for $\PH=\Phobs$ (inverted scenario)~\cite{Carena:2013ooa}.
In this Letter, the neutral \CP-even Higgs boson that is not identified as \Phobs is denoted \PGf.

Unsuppressed (or enhanced) quark couplings to BSM Higgs bosons are essential for many searches for extended Higgs sectors, such as that for the type-II 2HDM of the minimal supersymmetric SM (MSSM)~\cite{Branco:2011iw,Fayet:1974pd,Fayet:1977yc}. 
In these models, the dominant single Higgs boson production modes are gluon fusion and associated production with b quarks, that depend on the couplings to quarks.
However, in the case of type-X 2HDM, where lepton couplings are enhanced and up- and down-type quark couplings are suppressed by large values of \tanb, single Higgs boson production rates are diminished. 
This results in a parameter space that remains largely unexplored by existing LHC searches, which typically target ``MSSM-like'' models.

\begin{figure}[ht]
  \centering
  \includegraphics[width=\figurewidthfeynman]{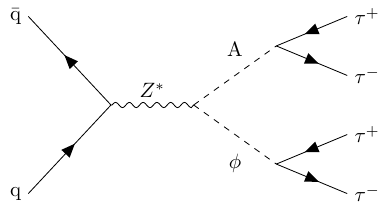}
  \caption{Feynman diagram for the production of two BSM neutral Higgs bosons from an off-shell \PZ boson and their decay to $\PGt$ leptons.}
  \label{fig:feynman}
\end{figure}

The type-X 2HDM has been proposed as a possible explanation for the discrepancy between the measured and predicted values of the muon anomalous magnetic moment~\cite{Carena:2002es,Ilisie:2015tra,Ghosh:2020tfq,Jueid:2021avn,Ferreira:2021gke,Abe:2015oca,Broggio:2014mna,Chun:2015hsa}.
The discrepancy is established from measurements of the muon anomalous magnetic moment with the SM prediction based on data-driven dispersive calculations~\cite{Muong-2:2006rrc,Muong-2:2021ojo,Muong-2:2023cdq,Aoyama:2020ynm}.
However, tensions between data-driven dispersive evaluations and lattice QCD calculations of the hadronic vacuum polarization contribution~\cite{Borsanyi:2020mff,Boccaletti:2024guq,Aliberti:2025beg} may reduce this discrepancy, and its interpretation as evidence for BSM physics remains uncertain.
Nevertheless, Ref.~\cite{Jueid:2021avn} identifies viable parameter space within a mass range accessible to searches at the LHC.
In this space, it is required that $\tanb \gtrsim 90$ in the normal scenario and $\tanb \gtrsim 120$ in the inverted scenario.
To search in this region, a production mode is required that is not suppressed by high-\tanb quark couplings.
This can be achieved by the simultaneous production of the \PGf and $\PA$ bosons from an off-shell \PZ boson, as shown in Fig.~\ref{fig:feynman}.
The cross section for this process is independent of \tanb and is predicted to vary between 10 and 600\unit{fb} at $\sqrt{s} = 13\TeV$ across the mass ranges $\mA \in [62.5, 145] \GeV$ and $\mphi \in [100, 245] \GeV$ found for the available parameter space in Ref.~\cite{Jueid:2021avn}.
At high \tanb in the type-X 2HDM, the branching fractions of BSM Higgs bosons to $\PGt$ leptons are dominant.

This region of interest lies beyond the sensitivity of single Higgs boson searches such as Ref.~\cite{CMS:2022goy}, which can only constrain the model at low values of \tanb due to suppressed production modes.
LHC di-Higgs searches, sensitive to heavy Higgs bosons, can be broadly categorized into three types: nonresonant standard model Higgs pair production ($H_{125}H_{125}$), heavy resonance decays to a Higgs and another scalar ($X \to YH_{125}$), and heavy resonance decays to two Higgs bosons ($X \to H_{125}H_{125}$). 
Searches targeting final states without $\PGt$ leptons are not compatible with the model under consideration, due to the dominant branching fraction of the additional Higgs bosons to $\PGt$ leptons. 
The remaining searches rely on the tagging of an SM-like Higgs boson, which removes the sensitivity to the signal model presented here. 
Consequently, none of the existing multi-$\PGt$ analyses target the unique topology of nonresonant $\PZst \to \PGf A \to 4\PGt$ production, in which both bosons decay exclusively to $\PGt$ leptons and no $\PH_{125}$ is present.
This search addresses a critical gap in current experimental coverage by leveraging the fact that the production cross section for $\PZst \to \PGf\PA$ is independent of \tanb, while the decay branching fractions to $\PGt$ leptons are enhanced at high \tanb.

This Letter reports the search for the $\PZst \to \PGf\PA \to \PGt\PGt\PGt\PGt$ process, using proton-proton collision ($\Pp\Pp$) events at $\sqrt{s} = 13\TeV$ collected with the CMS detector in 2016--2018, corresponding to an integrated luminosity of 138\fbinv~\cite{CMS-LUM-17-003,CMS-PAS-LUM-17-004,CMS-PAS-LUM-18-002}.
The search is sensitive to a wide range of masses of the \PGf and $\PA$ bosons that incorporate the regions of interest in the models proposed as an explanation for the muon anomalous magnetic moment measurement.
The results obtained are provided in the HEPData record for this analysis~\cite{hepdata}.

The CMS apparatus~\cite{CMS:2008xjf} is a multipurpose, nearly hermetic detector, designed to trigger on~\cite{CMS:2020cmk,CMS:2016ngn,CMS:2024aqx} and identify electrons ($\Pe$), muons ($\PGm$), photons, and (charged and neutral) hadrons~\cite{CMS:2020uim,CMS:2018rym,CMS:2014pgm}. 
A global ``particle-flow'' algorithm~\cite{CMS:2017yfk} aims to reconstruct all individual particles in an event, combining information provided by the all-silicon inner tracker and by the crystal electromagnetic (ECAL) and brass-scintillator hadron calorimeters, operating inside a 3.8\unit{T} superconducting solenoid, with data from the gas-ionization muon detectors embedded in the flux-return yoke outside the solenoid.
 
The reconstructed particles are used to build $\PGt$ leptons, jets, and missing transverse momentum (\ptvecmiss)~\cite{CMS:2018jrd,CMS:2016lmd,CMS:2019ctu}.
The \textsc{DeepTau} algorithm~\cite{CMS:2022prd} is used to further separate electrons, muons, and jets from hadronically decaying $\PGt$ leptons ($\tauh$).
The $\tauh$ discriminators against jets, electrons, and muons perform at $\tauh$ identification efficiencies of 80, 98, and 99.95\%, respectively.
Jets originating from the hadronization of $\PQb$ quarks ($\PQb$ jets) are tagged by the \textsc{DeepJet} algorithm~\cite{Sirunyan:2017ezt,Bols:2020bkb} at an approximate efficiency of 80\%.
Electrons are identified using a multivariate discriminant with a 90\% efficiency and a 1\% jet misidentification rate~\cite{CMS:2020uim}. 
Muons are identified using the quality of track fitting between the tracker and the muon system, as well as the consistency of individual track segments with the fitted track, achieving a 99\% efficiency with a 0.2\% pion misidentification rate~\cite{CMS:2018rym}.
Both electrons and muons are further required to be separated from any hadronic activity in the detector by requiring the relative isolation to be less than 0.15.
The relative isolation is calculated by adding up the transverse energy contributed by nearby particles and subtracting contributions from other pp vertices (``pileup''), within a cone of size $\Delta R = \sqrt{\smash[b]{(\Delta\eta)^2 + (\Delta\varphi)^2}} = 0.3\,(0.4)$ around the electron (muon) direction, and divided by the transverse momentum of the lepton, $\pt$.
Here $\Delta\eta$ and $\Delta\varphi$ are the angular distances in the pseudorapidity $\eta$ and azimuthal $\varphi$ (in radians) directions.

Six $\PGt\PGt\PGt\PGt$ final states are reconstructed in this search, accounting for $\approx$87\% of the branching fraction.
These are the $\tttt$, $\ettt$, $\mttt$, $\emtt$, $\mmtt$, and $\eett$ decay channels.
A $\ttt$ channel is also added to consider events in which all $\PGt$ leptons decay hadronically but one $\tauh$ candidate is lost due to reconstruction inefficiencies.
Events are selected by a trigger based on either the presence of a single electron, a single muon or a pair of $\tauh$ candidates.
In the different eras of data taking, the minimum $\pt$ requirement of the online trigger (off-line selection) lies in the range 25--32 (26--33) \GeV for an electron, 22--24 (23--25) \GeV for a muon and equals 35 (40) \GeV for a hadronic tau.
In channels with four objects in the final state, the sum of the charges of the object is required to be zero. 
In the $\ttt$ channel, the absolute value of the sum of the charges is required to be one.

The analysis is optimized separately in each decay channel to maximize signal sensitivity.
In the $\ettt$, $\mttt$, $\emtt$, $\mmtt$, and $\eett$ channels, events with a $\PQb$ jet are vetoed to remove a large portion of backgrounds from top quark pair production ($\ttbar$).
The $\eett$, $\mmtt$, and $\emtt$ channels are then split into two categories depending on whether the light leptons ($\Pe$ or $\PGm$) have same-sign (SS) or opposite-sign (OS) charges.
This helps separate the signal from the background from \ZLL or $\ttbar$ with two additional jets misidentified as $\tauh$ candidates ($\jtt$) and $\PZ(\to\Pell\Pell)\PZ(\to\tauh\tauh)$ production.
In these channels, the SS categories are more sensitive to the signal.

The final discriminating variable used is the total transverse mass ($\mttot$), defined as
  \begin{equation}
    \mttot = \sqrt{ \sum_{\substack{i,j=1; i<j}}^{N} \big[ \mt(\ptvec^{\hspace{2pt}i},\ptvec^{\hspace{2pt}j}) \big]^2 },  
  \end{equation}
where $\ptvec^{\hspace{2pt}i}$ is the vector of the $i^{\text{th}}$ element of an array containing the $\PGt$ decay objects (leptonic and hadronic) and \ptvecmiss, $N$ is the length of this array. The transverse mass $\mt$ is defined as,
  \begin{equation}
  \mt(\ptvec^{\hspace{2pt}i},\ptvec^{\hspace{2pt}j}) = \sqrt{2\pt^{i}\pt^{j}(1-\cos\Delta\varphi)},
  \label{eqn:mt}
  \end{equation}
where $\Delta\varphi$ is the azimuthal angle between $\ptvec^{\hspace{2pt}i}$ and $\ptvec^{\hspace{2pt}j}$.
The $\mttot$ shapes differ minimally between the different signal mass hypotheses.
To minimize large fluctuations in the background predictions, bin boundaries are set so that the statistical uncertainty in each bin is less than 25\% of the total.
If this condition cannot be fulfilled in a category, a single bin is used.

Three sources of background are considered: events with one or more \jtt candidate, events in which all \tauh candidates are correctly reconstructed, and a third component encompassing all other events that do not fit into these categories.
This final background is dominated by events in which a single light lepton is misidentified as a $\tauh$ candidate, while all other $\tauh$ candidates are correctly reconstructed.
Events with one or more \jtt candidate are modeled from data using an improved ``\FF method''~\cite{CMS:2018lkr} with machine learning (ML), while the remaining contributions are obtained from simulation. In many analysis categories, the number of predicted background events is small. 

The \FF method estimates the \jtt background in the signal regions by selecting events where \tauh candidates fail the nominal \textsc{DeepTau} identification (ID) requirement but pass a looser ID selection. 
Transfer functions (TFs) are then used to define event weights that are applied to correct the yields and kinematic distributions. 
The ML \FF method derives these TFs using boosted decision trees as in Ref.~\cite{Rogozhnikov:2016bdp}.
In this algorithm, decision trees partition the multidimensional feature space into regions where the source and target distributions differ.
The TFs are then updated over several iterations using an ensemble of trees, with each tree reducing the remaining differences between the two distributions.
This is done in a region of pure $\jtt$ objects, orthogonal to the signal region, by inverting one or more of the total charge selection or the ID requirements of other $\tauh$ candidates in the event.
The reweighting is performed with variables related to the \tauh candidate that are known to parametrize the TFs. 
These include the hadronic decay mode, \pt, the ratio \pt(jet)/\pt(\tauh) (where \pt(jet) refers to the \pt of the nearest jet within $\dR < 0.5$ of the \tauh), $\eta$, charge, and a boolean that indicates whether the object passed a leg of the \tauh pair trigger. 
The era of data taking and the $\pt$-ordered rank of the $\tauh$ candidate in the event are also included.
The TFs are applied to each $\tauh$ candidate in the event sequentially to estimate the background for events with one or more $\jtt$ candidate. 

The remaining backgrounds are modeled from simulation.
Backgrounds from diboson, triboson, \Wjets, \Zjets, \ttbar, single~\PQt quark, and $\Phobs\to\PZ\PZ$ production are considered.
The dominant contribution is from $\PZ\PZ$ events in which all $\tauh$ candidates are reconstructed correctly.
The \Wjets and \ZLL processes are simulated using the \MGvATNLO~2.4.2 event generator at leading order (LO) accuracy in the strong coupling constant \alpS~\cite{Alwall:2011uj,Alwall:2014hca}. 
Diboson and triboson production is simulated with \MGvATNLO~2.4.2 at next-to-LO (NLO) precision in \alpS~\cite{Frederix:2012ps,Alwall:2007fs}. 
The \ttbar~\cite{Alioli:2011as} and single~\PQt quark production samples~\cite{Frederix:2012dh} are generated at NLO precision using \POWHEG version 2.0~\cite{Nason:2004rx,Frixione:2007vw,Alioli:2010xd,Jezo:2015aia}. 
Their cross sections are calculated at next-to-NLO (NNLO) precision for \Wjets, \ZLL, \ttbar, and single~\PQt quark events in the $\PQt\PW$ channel~\cite{Melnikov:2006kv,Czakon:2011xx,Kidonakis:2013zqa}, and at NLO precision or higher for single~\PQt quark ($t$ channel) and diboson events~\cite{Kidonakis:2013zqa,Campbell:2011bn,Gehrmann:2014fva}.
Additionally, NNLO/NLO K factors, as derived in Ref.~\cite{CMS:2021ugl}, are applied to quark and gluon fusion initiated $\PZ\PZ$ production as a function of the $\PZ\PZ$ pair mass at the generator level.

The production of two BSM Higgs bosons from an off-shell \PZ boson in the alignment limit and its cross section are modeled in \MGvATNLO 2.6.5 at NLO precision in \alpS~\cite{Frederix:2012ps,Alwall:2007fs}.
The branching fractions of BSM Higgs bosons are calculated with \textsc{2HDECAY}~\cite{Krause:2018wmo}.
The alignment scenario is selected because any differences are minimal in signal generation or branching fractions out of the alignment scenario but within the experimental constraints from measurements of the \Phobs couplings.
Negligible changes to the signal modeling are observed when changing the free parameters $\mHc$ and $m_{12}^{2}$ within their theoretically allowed limits.

The \PYTHIA event generator~\cite{Sjostrand:2014zea} (version 8.230) is used to model parton showering and hadronization, as well as for \PGt lepton decays. 
The description of the underlying event is parametrized according to the CP5~\cite{Sirunyan:2019dfx} tune.
The NNPDF3.1~\cite{Ball:2017nwa} parton distribution functions (PDFs) are used for all simulations. 
For all simulated events, pileup collisions generated with \PYTHIA are added according to the expected pileup distribution in data. 
All events are passed through a \GEANTfour-based~\cite{Agostinelli:2002hh} simulation of the CMS detector and reconstructed with the same CMS event reconstruction software.

Several simulation-to-data corrections are applied to the simulation.
The corrections include adjustments for the trigger efficiency, tracking efficiency, energy scale (electrons only), identification and isolation requirements of the light lepton, and also for the trigger efficiency, identification and energy scale, calculated with the ``tag-and-probe'' scheme of the $\tauh$ candidate~\cite{Khachatryan:2010xn}. 
Corrections are also applied to account for discrepancies in pileup profiles, b tagging ID, \ptvecmiss, and jet energies. 
Finally, simulation-to-data corrections are used to account for differences in the misidentification rates of electrons and muons as tau candidates, as well as to match dilepton mass and $\pt$ spectra in Drell--Yan simulated events with data.
These corrections are performed in the same way as described in Ref.~\cite{CMS:2022goy}.

A 2\% uncertainty is applied to both the ID and trigger efficiencies of electrons and muons, which is treated as correlated between channels and applied to simulated processes.
The electron energy scale uncertainty is derived from the calibration of ECAL crystals.
Uncertainties ranging from 0--10\%, in the ID, energy scale, and trigger efficiencies of hadronic taus in simulated events are determined for separate decay modes and data-taking eras, and utilize a partial correlation scheme.
Uncertainties in jet energy calibration and resolution are applied with different correlations, varying from subpercent to around 10\%, depending on the kinematic observables of the jets. 
An uncertainty ranging from subpercent to 2\% is applied to the tagging of jets arising from gluons or light-flavor quarks rather than $\PQb$ jets.
An uncertainty is placed on the $\PZ\PZ$ differential cross section, equivalent to the size of the NNLO/NLO K factors.
All other theoretical uncertainties associated with the background modeling are found to be negligible and are, therefore, not considered.
Theoretical uncertainties in the signal from the renormalization and factorization scale, PDF sets, and $\alpha_S$ are considered.
A 1.6\% luminosity uncertainty is applied to all simulated samples~\cite{CMS-LUM-17-003,CMS-PAS-LUM-17-004,CMS-PAS-LUM-18-002}.
The statistical uncertainties in the background and signal templates are accounted for with the Barlow--Beeston lite procedure~\cite{Barlow:1993dm}.

The ML \FF method has a number of uncertainties.
An uncertainty in the nonclosure of the TFs, derived from the reweighted variables, is determined separately in each decay channel.
This has size $\mathcal{O}(10\%)$ on the total $\jtt$ background in each channel.
An uncertainty is determined to account for biases in the \FF TFs, resulting from inverting the ID and charge selections to obtain the pure $\jtt$ regions.
This uncertainty is uncorrelated across channels and is found to be around 5--15\%, depending on the channel.
Smaller uncertainties, that range from 0--5\% depending on the channel, are included to account for the influence of non-$\jtt$ objects in the measurement regions.

\begin{figure}[th]
  \centering
  \includegraphics[width=\figurewidthpostfit]{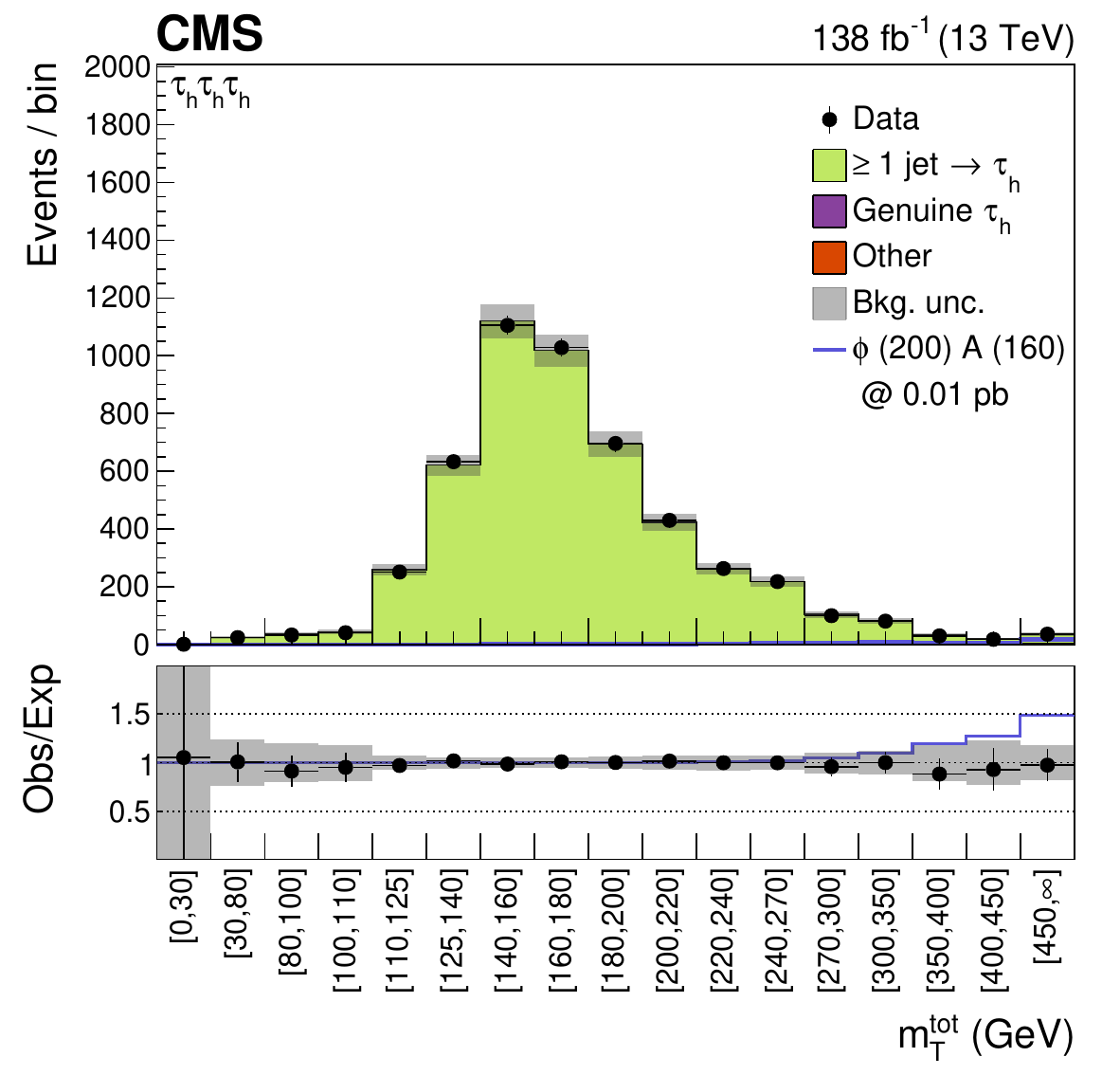}
  \includegraphics[width=\figurewidthpostfit]{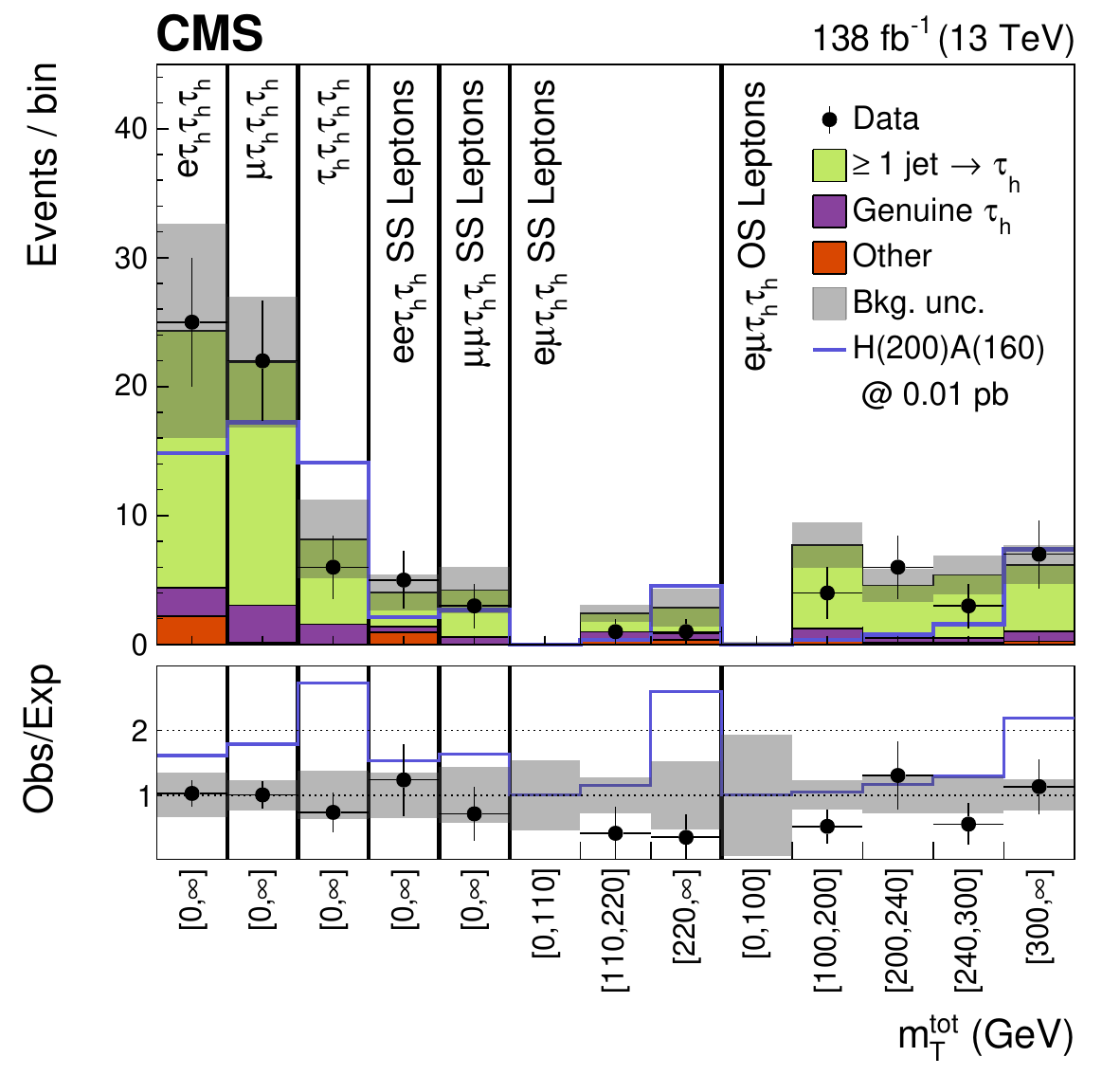}
  \caption{
  Distributions of the discriminating variable \mttot after a background-only fit to the data. 
  The $\ttt$ channel (\cmsLeft) and the $\ettt$, $\mttt$, and $\tttt$ channels, as well as the $\eett$ SS, $\mmtt$ SS, $\emtt$ SS, and $\emtt$ OS leptons categories are shown (\cmsRight).
  The background uncertainty represents the combined statistical and systematic uncertainties.
  An example signal with mass hypotheses $\mphi = 200$ and $\mA = 160\GeV$, with the cross section scaled to 0.01\unit{pb}, is shown as a blue line.
  }
  \label{fig:distributions}
\end{figure}

A profiled extended binned likelihood fit, where systematic uncertainties are treated as nuisance parameters, is used to infer the presence of a signal. 
This is performed using the CMS statistical analysis tool \textsc{Combine}~\cite{CMS:2024onh}.
The distribution of the expected background events and an example signal arbitrarily scaled, after a background-only fit to data, is shown in Fig.~\ref{fig:distributions} in the most sensitive analysis categories.
No significant deviation is observed between background prediction and the data.

The signal shape varies only slightly across the different mass hypotheses. 
For lower pair masses, the sensitivity is dominated by the $\mttt$ and $\tttt$ decay channels. 
As the pair mass increases, the $\mttot$ distribution shifts to higher values, making the greater statistical power and finer binning of the $\emtt$ and $\ttt$ channels increasingly important.

\begin{figure}[ht]
  \centering
  \includegraphics[width=\figurewidthmodelindependent]{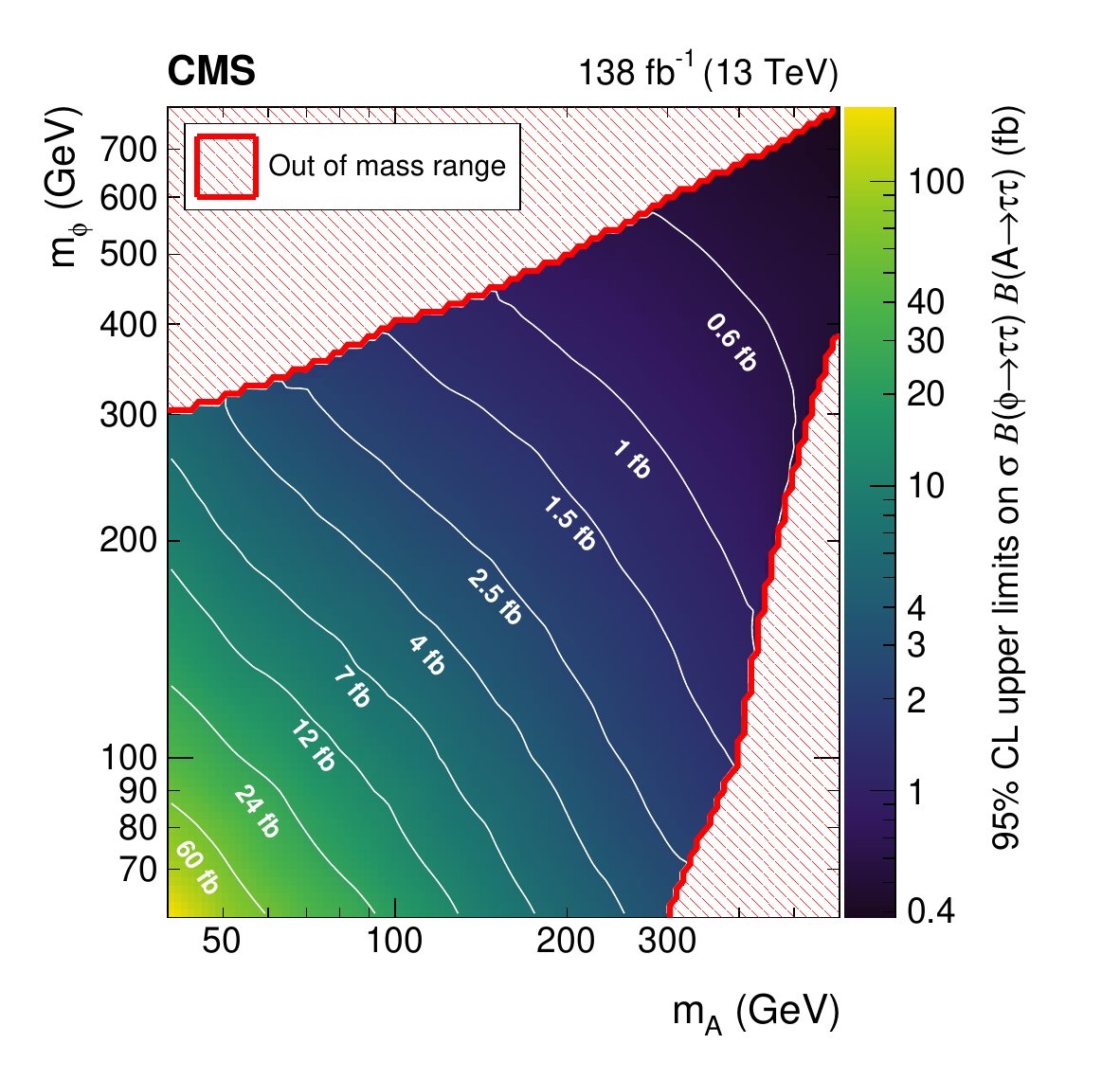}
  \caption{
    Observed 95\% $\CL$ upper limits on the product of the cross section ($\sigma$) for the production of two BSM Higgs bosons produced via an off-shell \PZ boson, and the branching fractions ($B$) for their decay into \PGt leptons.
    This is shown as a function of \mA and \mphi.
    No limits are set in the red hatched region. 
  }
  \label{fig:model_independent}
\end{figure}

Upper limits at 95\% confidence level ($\CL$) are determined using the asymptotic formulas for the distribution of test statistics~\cite{Cowan:2010js} and the modified frequentist approach~\cite{Junk:1999kv,Read:2002hq}.
First, upper limits are set on the production cross section times the branching fractions for every mass point tested and displayed in Fig.~\ref{fig:model_independent}.
The observed limits vary from 190\unit{fb} to 0.4\unit{fb}, for the mass hypotheses $\mA = 40$ and $\mphi = 60\GeV$ to $\mA = 600$ and $\mphi = 800\GeV$, respectively.

\begin{figure}[ht]
  \centering
  \includegraphics[width=\figurewidthmodeldependent]{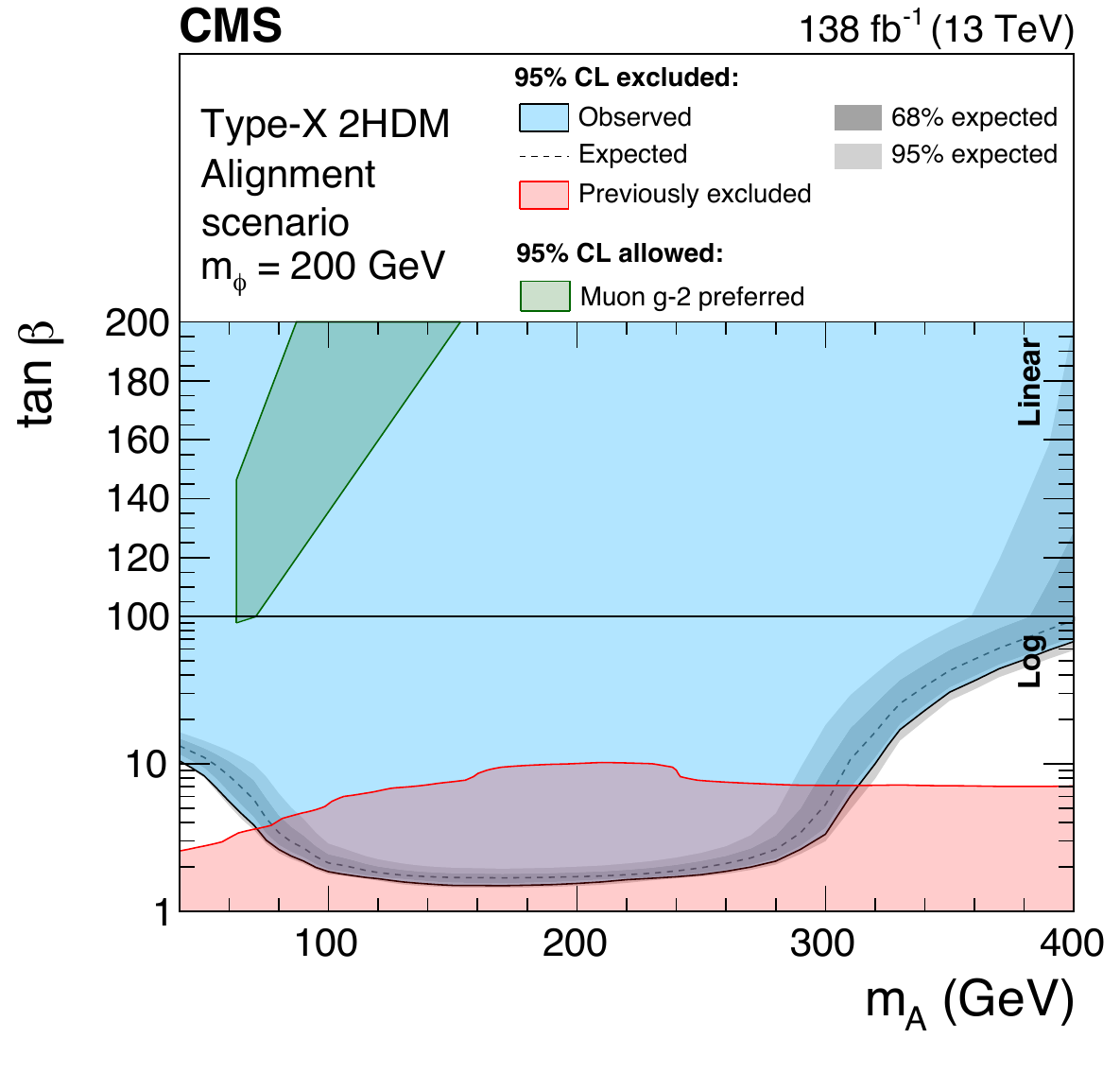}
  \caption{
   Observed and expected 95\% $\CL$ exclusion contours on the type-X 2HDM alignment scenario for $\mphi=200\GeV$.
   This is shown as a function of \mA and \tanb.
   The $y$-axis scales logarithmically below 100 and linearly above 100.
   The dashed line represents the expected median, the dark and bright grey bands correspond to the central 68 and 95\% expected intervals, and the blue area highlights the parameter space that has been excluded.
   Constraints from previous searches, obtained by \HIGGSTOOLS~\cite{Bahl:2022igd}, are shown in red.
   The allowed region in the normal alignment scenario accommodating the current measurement of the muon anomalous magnetic moment, as detailed in Ref.~\cite{Jueid:2021avn}, is shown in green.
  }
  \label{fig:model_dependent}
\end{figure}

An exclusion contour at 95\% $\CL$ in the \mA-\tanb plane for the type-X 2HDM alignment scenario with $\mphi=200\GeV$ is shown in Fig.~\ref{fig:model_dependent}.
In regions where \mA is larger than $\mphi-m_{\PZ}$ and \mphi is larger than $\mA-m_{\PZ}$, the exclusion limit does not depend strongly on \mA and reaches a \tanb value of approximately 1.5.
However, in regions where the separation of \mphi and \mA is greater than $m_{\PZ}$, the $\PH\to\PZ\PA$ or $\PA\to\PZ\PH$ decay becomes kinematically allowed and dominates the branching fraction.
In this case the limit is weaker.
The constraints determined from previous searches using \HIGGSTOOLS~\cite{Bahl:2022igd} are also shown in Fig.~\ref{fig:model_dependent}.
The strongest constraints are found from the CMS analysis of $\Ph/\PH/\PA \to \PGt\PGt$~\cite{CMS:2022goy}, that can constrain the parameter space to a minimum limit of $\tanb\approx 10$, which is approximately an order of magnitude smaller than the area of interest for the muon anomalous magnetic moment~\cite{Jueid:2021avn}.
This constraint cannot go higher in \tanb due to the suppressed single-Higgs boson production cross sections.
The limits are complementary to those of previous searches, as the production cross sections are independent of \tanb and can, therefore, exclude the previously allowed large values of \tanb.

\begin{figure}[th]
  \centering
  \includegraphics[width=\figurewidthmodeldependenttwod]{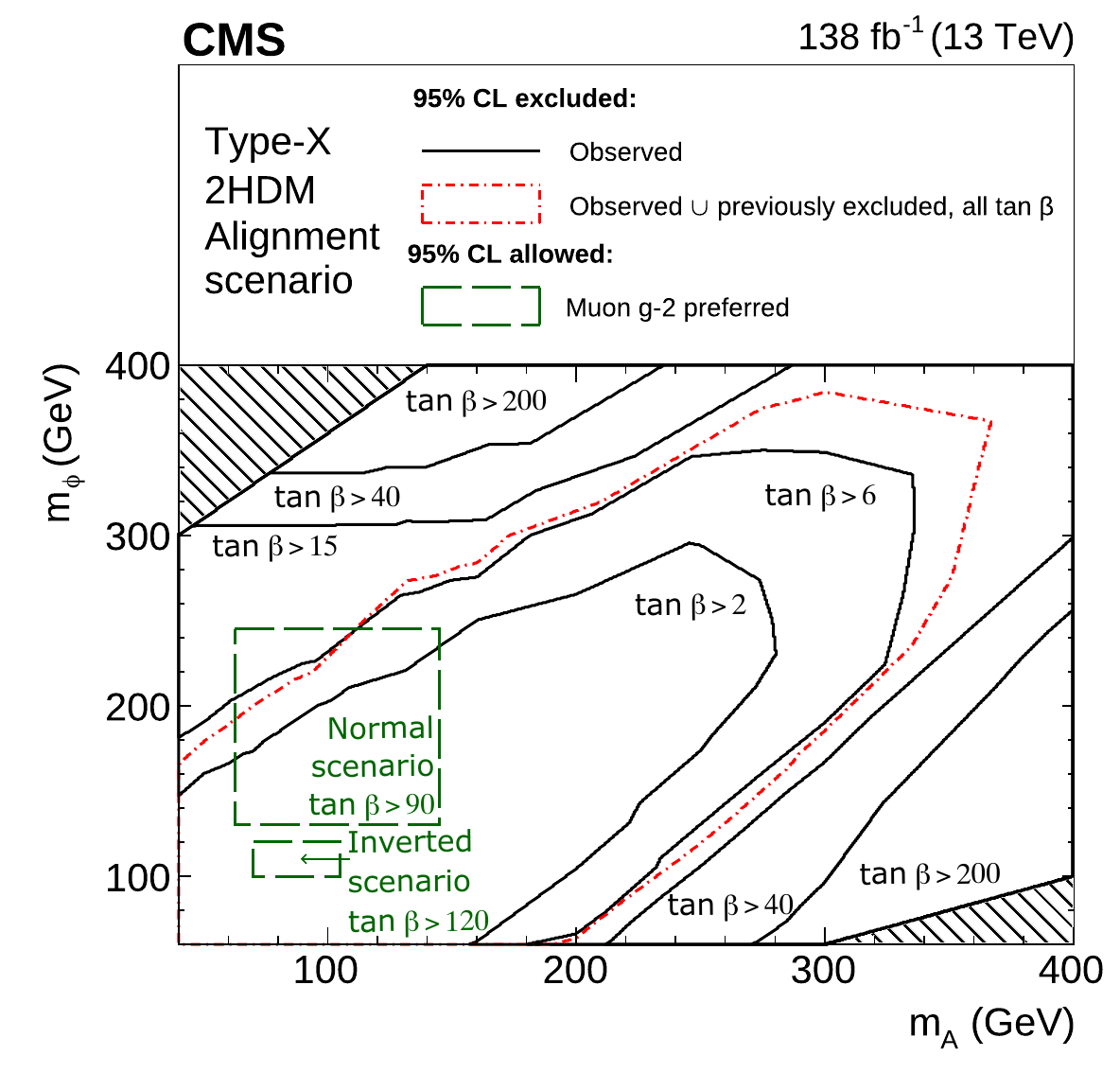}
  \caption{
   Observed 95\% $\CL$ exclusions of the type-X 2HDM alignment scenario in the $\mA$-$\mphi$ plane are shown as black solid contours, with the \tanb exclusion of each bounded region written on the plot.
   The complete exclusion of this model, from the union of the observed limits and those obtained by \HIGGSTOOLS~\cite{Bahl:2022igd}, is enclosed by the red dot-dashed contour.
   The allowed regions accommodating the current measurement of the muon anomalous magnetic moment, as detailed in Ref.~\cite{Jueid:2021avn}, are shown as green dashed boxes, with the \tanb allowed range written on the plot.
   These regions are excluded because they lie entirely within the parameter space for which $\tanb > 15$ is ruled out.
   Hatched regions indicate parameter space for which no exclusion limit is set.
  }
  \label{fig:model_dependent_2d}
\end{figure}

A summary of the exclusion contours in the \mA-\mphi plane are shown in Fig.~\ref{fig:model_dependent_2d}.
This also shows the complete exclusion contours of the type-X 2HDM alignment scenario from the union of the observed results and the limits obtained using \HIGGSTOOLS~\cite{Bahl:2022igd}, as well as the allowed regions for an explanation of the muon anomalous magnetic moment~\cite{Jueid:2021avn}.
Figure~\ref{fig:model_dependent_2d} demonstrates the exclusion of this model as an explanation for the muon anomalous magnetic moment, as the allowed regions at high \tanb in both the normal and inverted scenarios (green contours) are contained within the $\tanb > 15$ observed exclusion contour.
It also presents the complete exclusion of the type-X 2HDM alignment scenario down a diagonal in the \mA-\mphi plane (red contour), with an approximate width of 100\GeV, from the lowest mass point searched for at $\mA = 40$ and $\mphi = 60\GeV$ to $\mphi\approx\mA\approx360\GeV$.

In summary, a search has been presented for beyond-the-standard-model Higgs bosons produced via off-shell $\PZ$ bosons that decay into $\PGt$ leptons in pp collisions at the LHC, using a data sample collected with the CMS detector at $\sqrt{s} = 13\TeV$.
A good agreement between the background-only fit and the data is observed.
Upper limits on the production cross sections times the branching fractions range from 0.4\unit{fb} at $\mA = 600$ and $\mphi = 800\GeV$ to 190\unit{fb} at $\mA  = 40$ and $\mphi = 60\GeV$.
The data, combined with the constraints from previous searches, exclude a large portion of the phase space of the type-X two-Higgs-doublet model.
This search rules out this model as an explanation for the muon anomalous magnetic moment, as described in Ref.~\cite{Jueid:2021avn}.

\begin{acknowledgments}
We congratulate our colleagues in the CERN accelerator departments for the excellent performance of the LHC and thank the technical and administrative staffs at CERN and at other CMS institutes for their contributions to the success of the CMS effort. In addition, we gratefully acknowledge the computing centers and personnel of the Worldwide LHC Computing Grid and other centers for delivering so effectively the computing infrastructure essential to our analyses. Finally, we acknowledge the enduring support for the construction and operation of the LHC, the CMS detector, and the supporting computing infrastructure provided by the following funding agencies: SC (Armenia), BMBWF and FWF (Austria); FNRS and FWO (Belgium); CNPq, CAPES, FAPERJ, FAPERGS, and FAPESP (Brazil); MES and BNSF (Bulgaria); CERN; CAS, MoST, and NSFC (China); MINCIENCIAS (Colombia); MSES and CSF (Croatia); RIF (Cyprus); SENESCYT (Ecuador); ERC PRG and PSG, TARISTU24-TK10 and MoER TK202 (Estonia); Academy of Finland, MEC, and HIP (Finland); CEA and CNRS/IN2P3 (France); SRNSF (Georgia); BMFTR, DFG, and HGF (Germany); GSRI (Greece); MATE and NKFIH (Hungary); DAE and DST (India); IPM (Iran); SFI (Ireland); INFN (Italy); MSIT and NRF (Republic of Korea); MES (Latvia); LMTLT (Lithuania); MOE and UM (Malaysia); BUAP, CINVESTAV, CONACYT, LNS, SEP, and UASLP-FAI (Mexico); MOS (Montenegro); MBIE (New Zealand); PAEC (Pakistan); MSHE, NSC, and NAWA (Poland); FCT (Portugal); MESTD (Serbia); MICIU/AEI and PCTI (Spain); MOSTR (Sri Lanka); Swiss Funding Agencies (Switzerland); MST (Taipei); MHESI (Thailand); TUBITAK and TENMAK (T\"{u}rkiye); NASU (Ukraine); STFC (United Kingdom); DOE and NSF (USA).
\end{acknowledgments}

\bibliography{auto_generated}

\cleardoublepage \appendix\section{The CMS Collaboration \label{app:collab}}\begin{sloppypar}\hyphenpenalty=5000\widowpenalty=500\clubpenalty=5000\cmsinstitute{Yerevan Physics Institute, Yerevan, Armenia}
{\tolerance=6000
A.~Hayrapetyan, A.~Tumasyan\cmsAuthorMark{1}\cmsorcid{0009-0000-0684-6742}
\par}
\cmsinstitute{Institut f\"{u}r Hochenergiephysik, Vienna, Austria}
{\tolerance=6000
W.~Adam\cmsorcid{0000-0001-9099-4341}, J.W.~Andrejkovic, L.~Benato\cmsorcid{0000-0001-5135-7489}, T.~Bergauer\cmsorcid{0000-0002-5786-0293}, S.~Chatterjee\cmsorcid{0000-0003-2660-0349}, K.~Damanakis\cmsorcid{0000-0001-5389-2872}, M.~Dragicevic\cmsorcid{0000-0003-1967-6783}, P.S.~Hussain\cmsorcid{0000-0002-4825-5278}, M.~Jeitler\cmsAuthorMark{2}\cmsorcid{0000-0002-5141-9560}, N.~Krammer\cmsorcid{0000-0002-0548-0985}, A.~Li\cmsorcid{0000-0002-4547-116X}, D.~Liko\cmsorcid{0000-0002-3380-473X}, I.~Mikulec\cmsorcid{0000-0003-0385-2746}, J.~Schieck\cmsAuthorMark{2}\cmsorcid{0000-0002-1058-8093}, R.~Sch\"{o}fbeck\cmsAuthorMark{2}\cmsorcid{0000-0002-2332-8784}, D.~Schwarz\cmsorcid{0000-0002-3821-7331}, M.~Sonawane\cmsorcid{0000-0003-0510-7010}, W.~Waltenberger\cmsorcid{0000-0002-6215-7228}, C.-E.~Wulz\cmsAuthorMark{2}\cmsorcid{0000-0001-9226-5812}
% ...
\par}
\vskip\cmsinstskip
\dag:~Deceased\\
$^{1}$Also at Yerevan State University, Yerevan, Armenia\\
$^{2}$Also at TU Wien, Vienna, Austria\\
% ...
\end{sloppypar}
\end{document}